\documentclass[a4paper,11pt]{article}
\pdfoutput=1

\usepackage{jheppub}
\usepackage{graphicx}
\usepackage{amssymb}
\usepackage{cancel}
\usepackage{hyperref}
\usepackage{color}

\title{Improved bounds on $\mathbb{Z}_{3}$ singlet dark matter}

\renewcommand{\Re}{\operatorname{Re}}

\newcommand{\abs}[1]{\left\lvert #1 \right\rvert}

\author[a]{A. Hektor,}
\author[b,c]{A. Hryczuk,}
\author[a,1]{K. Kannike\note{Corresponding author.}}

\affiliation[a]{National Institute of Chemical Physics and Biophysics, \\
R\"{a}vala 10, Tallinn, Estonia}
\affiliation[b]{Department of Physics, University of Oslo, \\ Box 1048, NO-0316 Oslo, Norway}
\affiliation[c]{National Centre for Nuclear Research, \\ Ho\.{z}a 69, 00-681, 
Warsaw, Poland}

\emailAdd{Andi.Hektor@cern.ch}
\emailAdd{andrzej.hryczuk@ncbj.gov.pl}
\emailAdd{Kristjan.Kannike@cern.ch}

\abstract{We reconsider complex scalar singlet dark matter stabilised by a $\mathbb{Z}_{3}$ symmetry. We refine the stability bounds on the potential and use constraints from unitarity on scattering at finite energy to place a stronger lower limit on the direct detection cross section. In addition, we improve the treatment of the thermal freeze-out by including the evolution of the dark matter temperature and its feedback onto relic abundance. In the regions where the freeze-out is dominated by resonant or semi-annihilation, the dark matter decouples kinetically from the plasma very early, around the onset of the chemical decoupling. This results in a modification of the required coupling to the Higgs, which turns out to be at most few per cent in the semi-annihilation region, thus giving credence to  the standard approach to the relic density calculation in this regime. In contrast, for dark matter mass just below the Higgs resonance, the modification of the Higgs invisible width and direct and indirect detection signals can be up to a factor $6.7$. The model is then currently allowed at $56.8~\text{GeV}$ to $58.4~\text{GeV}$ (depending on the details of early kinetic decoupling) $\lesssim  M_{S} \lesssim 62.8~\text{GeV}$ and at $M_{S} \gtrsim 122~\text{GeV}$ if the freeze-out is dominated by semi-annihilation. We show that the whole large semi-annihilation region will be probed by the near-future measurements at the XENONnT experiment.}

\begin{document} 
\maketitle
\flushbottom

%%%%%%%%%%%%%%%%%%%%%%%%%%%%%%%%%%%%%%%%%%%%%%%%%%%%%%%%%%
\section{Introduction}

The discovery of the Higgs boson \cite{Chatrchyan:2012xdj,Aad:2012gk} has demonstrated that fundamental scalars do exist in Nature. The simplest addition to the Standard Model (SM) scalar sector that one can entertain is a scalar gauge singlet, which, stabilised by a $\mathbb{Z}_{2}$ symmetry, is a candidate of dark matter (DM) \cite{Silveira:1985rk,McDonald:1993ex,Burgess:2000yq} -- indeed one of the most popular and very well-studied DM candidates \cite{Davoudiasl:2004be,Ham:2004cf,OConnell:2006rsp,Patt:2006fw,Profumo:2007wc,Barger:2007im,He:2007tt,He:2008qm,Yaguna:2008hd,Lerner:2009xg,Farina:2009ez,Goudelis:2009zz,Profumo:2010kp,Guo:2010hq,Barger:2010mc,Arina:2010rb,Bandyopadhyay:2010cc,Drozd:2011aa,Djouadi:2011aa,Low:2011kp,Mambrini:2011ik,Espinosa:2011ax,Mambrini:2012ue,Djouadi:2012zc,Cheung:2012xb,Cline:2013gha,Urbano:2014hda,Endo:2014cca,Feng:2014vea,Duerr:2015bea,Duerr:2015mva,Beniwal:2015sdl,Cuoco:2016jqt,Escudero:2016gzx,Han:2016gyy,He:2016mls,Ko:2016xwd,Athron:2017kgt,Ghorbani:2018yfr}. 

Models of $\mathbb{Z}_{2}$ singlet DM are very predictive. The DM annihilation cross section, which determines the thermal relic abundance via the freeze-out mechanism, is set by the single Higgs portal coupling. The same coupling specifies the spin-independent direct detection cross section. For that reason, if a value for the DM mass is given, the relic density constraint directly determines the direct detection cross section. The direct detection experiments LUX \cite{Akerib:2016vxi}, PandaX-II \cite{Cui:2017nnn} and  XENON1T \cite{Aprile:2018dbl}, however, have not detected any sign of DM so far. These experiments have already ruled out $\mathbb{Z}_{2}$ singlet DM below the TeV scale, except in the narrow region around the Higgs resonance.

There are other symmetries, besides a $\mathbb{Z}_{2}$ group, that can be imposed on the scalar potential in order to stabilise DM. The next-to-simplest possibility is a $\mathbb{Z}_{3}$ symmetry that adds a cubic self-coupling term in the potential of the complex singlet. While the change may seem insignificant, in reality the DM phenomenology is considerably modified. The $\mathbb{Z}_{3}$ singlet scalar is the simplest dark matter candidate to undergo semi-annihilation \cite{Hambye:2008bq,Hambye:2009fg,Arina:2009uq,DEramo:2010ep}, which breaks the one-to-one correspondence between annihilation and direct detection cross section present in $\mathbb{Z}_{2}$-symmetric dark matter models. A significant semi-annihilation contribution to the relic density allows for a smaller Higgs portal and thus for a lower spin-independent direct detection cross section.

The $\mathbb{Z}_{3}$-symmetric complex singlet model was originally proposed in the context of neutrino physics \cite{Ma:2007gq}. First detailed analysis of DM phenomenology was carried out in \cite{Belanger:2012zr}. Similar behaviour of DM also occurs in a more complicated DM model based on $D_3$ symmetry~\cite{Adulpravitchai:2011ei}. Indirect detection of $\mathbb{Z}_{3}$ DM was considered in \cite{Arcadi:2017vis,Cai:2018imb}. Further developments in connection with radiative neutrino mass were studied in \cite{Aoki:2014cja,Bonilla:2016diq,Ding:2016wbd}. $\mathbb{Z}_{3}$ DM with several dark singlets was considered in \cite{Bhattacharya:2017fid}. The cubic coupling can contribute to $3 \to 2$ scattering for $\mathbb{Z}_{3}$ strongly interacting (SIMP) DM \cite{Choi:2015bya,Choi:2016tkj}.

It is interesting to note that the $\mathbb{Z}_{3}$ symmetry can be the remnant of a dark $U(1)$ local \cite{Ko:2014loa,Ko:2014nha,Guo:2015lxa} or global \cite{Bernal:2015bla} symmetry. Alternatively, the $\mathbb{Z}_{3}$ singlet DM can be considered to be a limiting case of more complicated models that also include an inert doublet \cite{Kadastik:2009cu,Kadastik:2009dj,Belanger:2012vp,Belanger:2014bga,Bonilla:2014xba}, which have quartic semi-annihilation couplings and can have two-component dark matter. Semi-annihilation and multi-particle dark matter can arise in multi-Higgs-doublet models with $\mathbb{Z}_{N}$ symmetry \cite{Ivanov:2012hc} or models with a dark $SU(N)$ symmetry \cite{Karam:2016rsz,Karam:2015jta} as well.

A usual assumption of the standard calculation of the thermal relic abundance \cite{Gondolo:1990dk} is that at the time of freeze-out DM is still in local equilibrium with the heat bath. This is motivated by the elastic scattering processes with the thermal bath particles typically being much more efficient than the annihilation and production processes. However, if the latter are enhanced (e.g., by a resonance), or in general when scattering processes are unrelated to the number changing ones (as is the case of semi-annihilation) there is no reason to expect that this standard assumption is satisfied. Indeed, it has been shown recently \cite{Binder:2017rgn} (see also \cite{Duch:2017nbe}) that kinetic decoupling can get under way as early as the chemical one and the subsequent change of the shape of the DM phase space distribution can  modify the relic abundance by even more than an order of magnitude.
In fact, the concrete example that is given in \cite{Binder:2017rgn} is the case of $\mathbb{Z}_{2}$ scalar singlet dark matter around the Higgs resonance, which finds its clear analogue in the Higgs resonance region of the $\mathbb{Z}_{3}$ singlet DM. 

The model studied in this work has another open region, where the Higgs portal coupling is suppressed, i.e. when the DM relic density is mainly determined by semi-annihilation, which additionally results in self-heating of the DM component. This raises the question whether the DM kinetic decoupling also happens early in this case and whether this alters the resulting relic abundance.

The strength of semi-annihilation depends on the strength of the cubic self-coupling of the singlet. The coupling is bounded by the requirement that the electroweak (EW) minimum be the global one. Lately, it was pointed out that perturbative unitarity for scattering at finite energy can put bounds on dimensionful couplings of new physics models \cite{Goodsell:2018tti}. 

A global likelihood fit of the $\mathbb{Z}_{3}$ singlet dark matter was recently made with the GAMBIT code \cite{Athron:2018ipf}. This study did not, however, take into account the improvements included in this work, in particular refined unitarity bounds and treatment of early kinetic decoupling. These developments are especially relevant precisely in the regions that are still allowed by the experimental data and where the improved precision of theoretical predictions is required for robust claims of exclusion of the whole parameter space of the thermal $\mathbb{Z}_{3}$ singlet dark matter model.

The aim of this paper is to provide a timely update of the past results \cite{Belanger:2012zr}. While the unitarity constraints are often computed in the limit of infinite energy, we calculate them at finite energy with the help of the latest version \cite{Goodsell:2018tti} of the SARAH package \cite{Staub:2009bi,Staub:2010jh,Staub:2012pb,Staub:2013tta}. We use the one-loop effective potential to calculate the bounds of absolute stability and metastability of the EW minimum from the tunnelling rate with the help of the AnyBubble package \cite{Masoumi:2016wot}.\footnote{The first-order phase transition from thermal tunnelling \emph{into} the EW minimum can produce a measurable gravitational wave signal, but only in a parameter space region with DM underdensity \cite{Kang:2017mkl}.} These constraints, in particular the one from the unitarity, put an upper bound on the singlet cubic self-coupling and therefore on the semi-annihilation cross section. We take into account early kinetic decoupling around the Higgs resonance and for large semi-annihilation, and use the micrOMEGAs code \cite{Belanger:2018mqt} to calculate relic density in the larger part of the parameter space. The micrOMEGAs is also used to compute predictions for direct and indirect detection signals. A large part of the parameter space is already ruled out by XENON1T \cite{Aprile:2018dbl}. Thanks to the new unitarity constraints, we manage to further restrict the model.

We introduce the model in section~\ref{sec:model}. Various theoretical and experimental constraints are considered in section~\ref{sec:constraints}. Dark matter freeze-out, the impact of early kinetic decoupling and semi-annihilation are studied in section~\ref{sec:relic}. Section~\ref{sec:indirect} discusses prospects of direct and indirect detection of dark matter. We conclude in section~\ref{sec:conclusions}. Details of the field-dependent masses and counter-terms for the effective potential are given in the appendix~\ref{sec:counter}.

%%%%%%%%%%%%%%%%%%%%%%%%%%%%%%%%%%%%%%%%%%%%%%%%%%%%%%%%%%
\section{The model}
\label{sec:model}

The most general renormalisable scalar potential of the Higgs doublet $H$ and the complex singlet $S$, invariant under the $\mathbb{Z}_3$ transformation $H \to H$, $S \to e^{i 2 \pi/3} S$, is given by
\begin{equation}	
  V = \mu_{H}^{2} \abs{H}^{2} + \lambda_{H} \abs{H}^{4} 
  + \mu_{S}^{2} \abs{S}^{2} + \lambda_{S} \abs{S}^{4}
  + \lambda_{SH} \abs{S}^{2} \abs{H}^{2} + \frac{\mu_3}{2} (S^{3} + S^{\dagger 3}).
\label{eq:V:Z:3:singlet}
\end{equation}
This is the only possible potential with this field content and symmetry. Without loss of generality, we can take $\mu_{3}$ real and non-negative. 

The mass of the Higgs boson is $M_{h} = 125.09$~GeV \cite{Aad:2015zhl} and the Higgs vacuum expectation value (VEV) $v = 246.22$~GeV. We fix the parameters
\begin{equation}
\begin{split}
  \mu_H^{2} &= -\frac{M_h^{2}}{2}, \\
  \lambda_H &= \frac{1}{2} \frac{M_h^{2}}{v^2}, \\ 
  \mu_S^2 &= M_S^2 - \lambda_{SH} \frac{v^2}{2}.
\end{split}
\label{eq:parameters}
\end{equation}
Dark matter mass $M_{S}$, the Higgs portal $\lambda_{SH}$, the singlet cubic coupling $\mu_{3}$ and the singlet quartic self-coupling $\lambda_{S}$ are left as free parameters.

%%%%%%%%%%%%%%%%%%%%%%%%%%%%%%%%%%%%%%%%%%%%%%%%%%%%%%%%%%
\section{Theoretical and experimental constraints}
\label{sec:constraints}

%%%%%%%%%%%%%%%%%%%%%%%%%%%%%%%%%%%%%%%%%%%%%%%%%%%%%%%%%%
\subsection{Vacuum stability}
\label{sec:vac:stab}

In order to ensure a finite minimum for the potential energy, the scalar potential must be bounded from below in the limit of larger field values, in which case dimensionful terms can be neglected. The potential \eqref{eq:V:Z:3:singlet} is bounded from below if
\begin{equation}
  \lambda_{H} > 0, \quad \lambda_{S} > 0, \quad \lambda_{SH} + 2 \sqrt{\lambda_{H} \lambda_{S}} > 0.
\end{equation}

%%%%%%%%%%%%%%%%%%%%%%%%%%%%%%%%%%%%%%%%%%%%%%%%%%%%%%%%%%
\subsection{Perturbativity}
\label{sec:pert}

To ensure validity of perturbation theory, loop corrections to couplings should be smaller than their tree-level values. The model is perturbative \cite{Lerner:2009xg} if $\abs{\lambda_{S}} \leqslant \pi$ and $\abs{\lambda_{SH}} \leqslant 4 \pi$.

%%%%%%%%%%%%%%%%%%%%%%%%%%%%%%%%%%%%%%%%%%%%%%%%%%%%%%%%%%
\subsection{Unitarity}
\label{sec:unit}

Perturbative unitarity constraints arise from the unitarity of the $S$-matrix for the $2 \to 2$ scalar field scattering amplitudes. At the order of the zeroth partial wave, the matrix is given by
\begin{equation}
  a_{0}^{ba} = \frac{1}{32 \pi} \sqrt{\frac{4 \abs{\mathbf{p}^{b}} \abs{\mathbf{p}^{a}}}{2^{\delta_{a}} 2^{\delta_{b}} s}} \int_{-1}^{1} d (\cos \theta) \mathcal{M}_{ba}(\cos \theta),
 \end{equation}
 where a pair $a$ of scalars scatters to another pair $b$ with the matrix element $\mathcal{M}_{ba}(\cos \theta)$, where $\theta$ is the angle between the incoming ($ \abs{\mathbf{p}^{a}}$) and outgoing ($ \abs{\mathbf{p}^{b}}$) three-momenta in the centre-of-mass frame, and $s = (p_{1} + p_{2})^{2}$ is a Mandelstam variable. The exponent $\delta_{a}$ is unity if the particles in pair $a$ are identical and zero otherwise; similarly for $\delta_{b}$ and pair $b$. The eigenvalues $a_{0}^{i}$ of the scattering matrix must satisfy
\begin{equation}
  \abs{\Re a_{0}^{i}} \leqslant \frac{1}{2}.
  \label{eq:unit:def}
\end{equation}

It is usual to calculate unitarity constraints only in the limit of infinite scattering energy $s \to \infty$, in which case only quartic couplings contribute to scattering. Having said that, the full calculation at finite energy that includes all tree-level contributions can set more stringent constraints, in particular on trilinear couplings \cite{Schuessler:2007av}.

We have implemented the model for the SARAH package \cite{Staub:2013tta} and used it to calculate unitarity constraints at finite scattering energy \cite{Goodsell:2018tti}. The $S$-matrix also takes into account scattering of the Goldstone bosons $G^{0}$ and $G^{\pm}$. The limits are calculated in the Feynman gauge, where their masses are $M_{G^{0}} = M_{Z}$ and $M_{G^{\pm}} = M_{W^{\pm}}$.

The unitarity bounds \eqref{eq:unit:def} in the $s \to \infty$ limit are given by
\begin{align}
  \abs{\lambda_{H}} \leqslant 4 \pi,  \quad\qquad \abs{\lambda_{S}} \leqslant 4 \pi, \quad\qquad  \abs{\lambda_{SH}} &\leqslant 8 \pi,
  \\
  \abs{3 \lambda_{H} + 2 \lambda_{S} \pm \sqrt{9 \lambda_{H}^2 - 12 \lambda_{H} \lambda_{S} + 4 \lambda_{S}^2 + 2 \lambda_{SH}^{2}}} &\leqslant 8 \pi,
\end{align}
where the last condition, in the $\lambda_{SH} = 0$ limit, yields $\abs{\lambda_{H}} \leqslant \frac{4}{3} \pi$ and $\abs{\lambda_{S}} \leqslant 2 \pi$.

At finite energy, we take $s \geqslant 4 M_{S}^{2}$ to avoid spurious poles \cite{Schuessler:2007av}. The eigenvalues of the scattering matrix cannot be given analytically. An approximation can be obtained, however, if we set $ \lambda_{H} = \lambda_{SH} = 0$. Then the non-zero eigenvalues of the $S$-matrix are
\begin{align}
  a_{0}^{1} &= -\frac{\sqrt{s (s-4 M_{S}^{2})} 
  \left[ 4 \lambda_{S} (s -M_{S}^{2}) + 9 \mu_{3}^{2} \right]}%
  {32 \pi s (s-M_{S}^{2})},
   \\
   a_{0}^{2} &=  \frac{4 \lambda_{S} (4 M_{S}^{2} - s) 
   + 9 \mu_{3}^{2} \ln \frac{s-3 M_{S}^{2}}{M_{S}^{2}}}{16 \pi  \sqrt{s (s-4 M_{S}^{2})}}.
\end{align}
It is $a_{0}^{2}$ that gives a stronger limit with a maximum at about $s \approx 5 M_{S}^{2}$. 

%%%%%%%%%%%%%%%%%%%%%%%%%%%%%%%%%%%%%%%%%%%%%%%%%%%%%%%%%%
\subsection{Stability of the electroweak vacuum}
\label{sec:tunnel}

The quantum corrections to the potential in the $\overline{\text{MS}}$ renormalisation scheme are given, at one-loop level, by
\begin{equation}
  \Delta V = \sum_{i} \frac{1}{64 \pi^{2}} n_{i} m^{4}_{i} \left( \ln \frac{m_{i}^{2}}{\mu^{2}} - c_{i} \right),
\end{equation}
where $n_{i}$ are the degrees of freedom of the $i$th field, $m_{i}$ are field-dependent masses and the constants $c_{i} = \frac{3}{2}$ for scalars and fermions and $c_{i} = \frac{5}{2}$ for vector bosons. To calculate the effective potential in case of negative field-dependent masses, we substitute $\ln m_{i}^{2} \to \ln \abs{m_{i}^{2}}$, which is equivalent to analytical continuation \cite{Bobrowski:2014dla}.
We add a counter-term potential
\begin{equation}	
\begin{split}
  \delta V &= \delta\mu_{H}^{2} \abs{H}^{2} + \delta\lambda_{H} \abs{H}^{4} 
  + \delta\mu_{S}^{2} \abs{S}^{2} + \delta\lambda_{S} \abs{S}^{4}
  + \delta\lambda_{SH} \abs{S}^{2} \abs{H}^{2} 
  \\
  &+ \frac{\delta\mu_3}{2} (S^{3} + S^{\dagger 3}) + \delta V_{0},
\end{split}
\label{eq:delta:V:Z:3:singlet}
\end{equation}
chosen as to retain some of the properties of the tree-level potential: positions of minima, masses in the electroweak minimum, and the size of the cubic coupling $\mu_{3}$. 
The effective potential is then
\begin{equation}
  V^{(1)} = V + \Delta V + \delta V.
\end{equation}
Details on the field-dependent masses and counter-terms are given in appendix~\ref{sec:counter}. We pick the renormalisation scale $\mu = M_{t}$.

The one-loop level absolute stability bound is relaxed in comparison to the tree-level absolute stability bound in the limit of small $\lambda_{SH}$,
\begin{equation}
  \max \frac{\mu_3}{M_S} \approx 2 \sqrt{\lambda_S}.
  \label{eq:abs:stab:tree:level}
\end{equation}
The tree-level and one-loop results differ by up to 5\% for perturbative values of $\lambda_{S}$. For $\lambda_{SH} < 0$ the constraints are rather stricter than for $\lambda_{SH} \gtrsim 0$, but with regard to relic density, we can choose the latter without loss of generality.

In order to determine the metastability bound, we calculate the tunnelling rate with the help of the AnyBubble code \cite{Masoumi:2016wot}. In practice, the approximation made in the previous paper \cite{Belanger:2012zr} that considers only tunnelling in the $s$-coordinate and uses a numerical fit \cite{Adams:1993zs} to the Euclidean action holds up quite well. The reason is that tunnelling between vacua with opposite sign of the Higgs VEV is much harder than tunnelling between same-sign vacua \cite{Branchina:2018qlf}. Indeed, one can use the approximate bound \eqref{eq:abs:stab:tree:level} as a seed value for the calculation of maximal allowed value of $\mu_{3}$.

The bubble nucleation rate per unit time and volume is given by
\begin{equation}
  \Gamma = D e^{-S_{E}} \simeq \phi_{0}^{4} e^{-S_{E}},
\end{equation}
where $\phi_{0}$ is the field value at the center of the bounce and $S_{E}$ is the Euclidean action. Metastability of the vacuum means that not one bubble has nucleated within the past light-cone with volume $V$ and lifetime $T$ of the Universe:
\begin{equation}
  \Gamma V T \approx \Gamma H_{0}^{-4} < 1,
\end{equation}
where $H_{0}$ is the Hubble constant.

The metastability bound is only a few per cent larger than the absolute stability bound. Note that even if the vacuum is metastable at zero temperature, it may be possible that the Universe can end up in the deeper minima in the first place via thermal tunnelling, further reducing the parameter space \cite{Camargo-Molina:2014pwa}.

%%%%%%%%%%%%%%%%%%%%%%%%%%%%%%%%%%%%%%%%%%%%%%%%%%%%%%%%%%
\subsection{Higgs invisible width}
\label{sec:invis}

The decay width of the Higgs boson to singlets is
\begin{equation}
  \Gamma_{h \to S S^{*}}^{\mathbb{Z}_{3}} = \frac{\lambda_{SH}^{2} v^{2}}{16 \pi M_{h}} \sqrt{1 - \frac{4 M_{S}^{2}}{M_{h}^{2}}}
\end{equation}
for $M_{S} \leqslant M_{h}/2$.
In the Standard Model, the Higgs total decay width is $\Gamma_{h}^{\text{SM}} = 4.07 \times 10^{-3}$~GeV \cite{PhysRevD.98.030001}. The invisible branching ratio $\text{BR}_{\text{inv}} = \Gamma_{h \to S S^{*}}^{\mathbb{Z}_{3}}/\Gamma_{h}^{\text{SM}}$ is constrained to be below about $0.24$ at $95\%$ confidence level \cite{Khachatryan:2016whc,ATLAS-CONF-2018-031} by direct measurements and below about $0.17$ by statistical fits of all Higgs couplings \cite{Giardino:2013bma,Belanger:2013xza}.

%%%%%%%%%%%%%%%%%%%%%%%%%%%%%%%%%%%%%%%%%%%%%%%%%%%%%%%%%%
\subsection{Combination of constraints}
\label{sec:constraint:combo}

\begin{figure}[tb]
\begin{center}
  \includegraphics{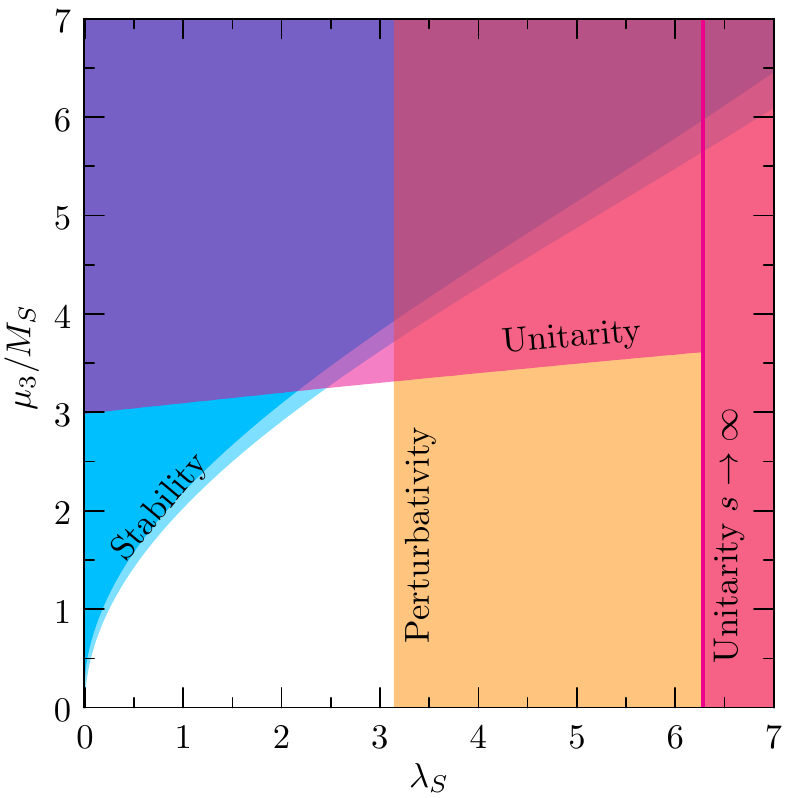}
\caption{Constraints on the parameter space. In the blue area, the EW-breaking, $\mathbb{Z}_{3}$-symmetric vacuum is not stable (is metastable for light blue). Yellow marks lack of perturbativity and magenta violation of perturbative unitarity. The vertical magenta line marks the bound on unitarity in the limit $s \to \infty$. The white area is allowed by all constraints.}
\label{fig:bounds}
\end{center}
\end{figure}

The combined bounds from vacuum stability, perturbativity and unitarity are shown in figure~\ref{fig:bounds}. The bounds are given for the Higgs portal $\lambda_{SH} \approx 0$, which is a good approximation for the parameter space determined by the relic density. The yellow area is forbidden by the perturbativity bound $\abs{\lambda_{S}} \leqslant \pi$, above which the one-loop and tree-level contributions from the scalar self-coupling $\lambda_{S}$ become comparable in size. The perturbative unitarity constraints from scattering at finite energy are shown in magenta. The usual unitarity constraint in the infinite energy limit $s \to \infty$ produces $\abs{\lambda_{S}} \leqslant 2 \pi$, which is given by the vertical magenta line. In both cases, the dependence of the unitarity bound on the Higgs portal remains at per cent level even up to $\lambda_{SH} \simeq 1$. In the blue area, the absolute stability of our vacuum is violated as the $\mathbb{Z}_{3}$-breaking minimum becomes lower in energy. In the light blue area the universe is metastable; the dependence of these bounds on $\lambda_{SH}$ is negligible in the parameter range determined by the relic density. In the region of high $\lambda_{S}$, it is the unitarity and perturbativity bounds that constrain the parameter space, while for lower $\lambda_{S}$, it is the stability bound. The white area is allowed by all constraints.

%%%%%%%%%%%%%%%%%%%%%%%%%%%%%%%%%%%%%%%%%%%%%%%%%%%%%%%%%%
\section{Relic density}
\label{sec:relic}

The requirement that the thermal relic density be equal to the observed value $\Omega_{c} h^{2} = 0.120 \pm 0.001$ \cite{Aghanim:2018eyx} provides a very strong constraint on the model. For an accurate determination of the theoretical prediction matching the precision of measurements, we use the treatment first introduced in \cite{vandenAarssen:2012ag} and extended to the early kinetic decoupling regime in \cite{Duch:2017nbe,Binder:2017rgn}, based on solving the coupled system of Boltzmann equations (cBE) for the number density and the second moment of the phase space distribution.\footnote{In \cite{Binder:2017rgn} an even more general method was discussed, based on the full numerical solution for the phase space distribution function of DM, allowing to accurately treat the impact of any possible deviations from the Maxwell-Boltzmann shape. It has been shown, however, that such a detailed approach is not necessary when there are efficient DM self-interactions or when the velocity dependence of the annihilation process is not very strong -- as it would be e.g. in the case of the $s$-channel resonance -- and that then the coupled system of Boltzmann equations provides a very good approximation.}

Whenever the kinetic decoupling happens significantly later than chemical one we use the micrOMEGAs 5.0 \cite{Belanger:2018mqt} to calculate relic density. Then the kinetic equilibrium is enforced during freeze-out and the equations for the number density, $n$, take the form \cite{Belanger:2012vp}
\begin{equation}
\frac{dn}{dt}+3Hn=-v\sigma^{S S^* \rightarrow  XX}  \left(n^2-\overline{n}^2 \right) -\frac{1}{2} v\sigma^{SS\rightarrow
S^* h}  
\left(n^2-n \, \overline{n} \right),
\end{equation} 
where $X$ is any SM particle. There is only one possible semi-annihilation process $S S \to S^* h$ in this model  that goes through $S$ exchange in $s$-, $t$- and $u$-channels. For kinematical reasons, this process is allowed only for $\sqrt{s} > M_S+M_h$. 

In regimes close to the Higgs resonance and when semi-annihilation becomes relevant, we solve the cBE for both the zeroth and second moment of the DM phase space distribution $f$, defined as
\begin{equation}
Y\equiv \frac{n}{s}=\frac{g}{s}\int \frac{d^3p}{(2\pi)^3}\,f(\mathbf{p}), \qquad y\equiv 
\frac{M_{\rm DM}}{3n s^{2/3}}
g\int \frac{d^3p}{(2\pi)^3}\,\frac{\mathbf{p}^2}{E} f(\mathbf{p}),
\end{equation}
where $g$ is the number of internal degrees of freedom of the DM particle and $s$ is the entropy density. The parameter $y$ can be used to define the DM temperature through $T_{\rm DM}=ys^{2/3}/m_{\rm DM}$. The coupled system takes the form \cite{Binder:2016pnr}
\begin{eqnarray}
\frac{Y'}{Y} &=&  \frac{g}{x \tilde H n} \int \frac{d^3p}{(2\pi)^3E}\, C[f]\,, \label{Yfinal}\\
\frac{y'}{y} &=& \frac{g}{3x \tilde Hn T_{\rm DM}} \int \frac{d^3p}{(2\pi)^3E} \frac{\mathbf{p}^2}{E}\, C[f] - \frac{Y'}{Y} 
+\frac{H}{\tilde H}
\frac{g}{3xnT_{\rm DM}}\int \frac{d^3p}{(2\pi)^3}\,\frac{\mathbf{p}^4}{E^3} f(\mathbf{p})\,,\label{yfinal}
\end{eqnarray}
where $'\equiv d/dx$, $x=M_{\rm DM}/T$ and $\tilde H\equiv H/\left(1+  \frac{1}{3} \frac{T}{g^s_{\rm\text{eff}}}\frac{ d g^s_{\rm\text{eff}}}{d T}\right)$. The collision term $C[f]$ contains contributions from all possible interactions including elastic scattering, annihilation and semi-annihilation. The explicit expressions for the collision term and its moments can be found in  \cite{Binder:2017rgn} and for the contribution of semi-annihilation in the appendix A of \cite{Cai:2018imb}.\footnote{The cBE system appropriate for semi-annihilation has been first derived in \cite{Kamada:2018hte} and used in the context of halo core formation and then studied also in \cite{Cai:2018imb} with applications to dark matter indirect detection.  The main difference between these previous implementations and the one used in this work is that for the relic density calculation one cannot assume a deeply non-relativistic regime. This requires more computationally expensive numerical evaluations of the up-to-three-dimensional phase space integrals to determine the thermal averages, which additionally depend not only on $T$ but also on $T_{\rm DM}$.} 

This system of cBE has been then used to find the improved relic density for the points situated on the boundary of the parameter space set by the absolute stability bound, i.e. points that give the correct relic abundance using standard approach and have the lowest allowed value of $\lambda_{SH}$. In order to be conservative, the computation was done assuming the smallest scattering scenario for QCD phase transition (i.e. case `B' discussed in \cite{Binder:2017rgn}) where only the light quarks ($u$, $d$, $s$) contribute to the scattering, and only above hadronisation temperature taken to be $600$~MeV.\footnote{For the Higgs resonance region shown in figure~\ref{fig:direct:detection} below, in order to bracket the uncertainty related to QCD phase transition, the results are shown for both this case and also the largest scattering scenario. The latter assumes that all quarks are free and present in the plasma down to critical temperature of $T_{c} = 154$~MeV (i.e. case `A' discussed in \cite{Binder:2017rgn}).}
It was found that in the semi-annihilation regime of the  $\mathbb{Z}_{3}$ singlet DM model the modification of the final relic density is at the level of at most few per cent.
This warrants to use the micrOMEGAs code for determination of the relic density calculation in that regime. 

\begin{figure}[tb]
\begin{center}
  \includegraphics{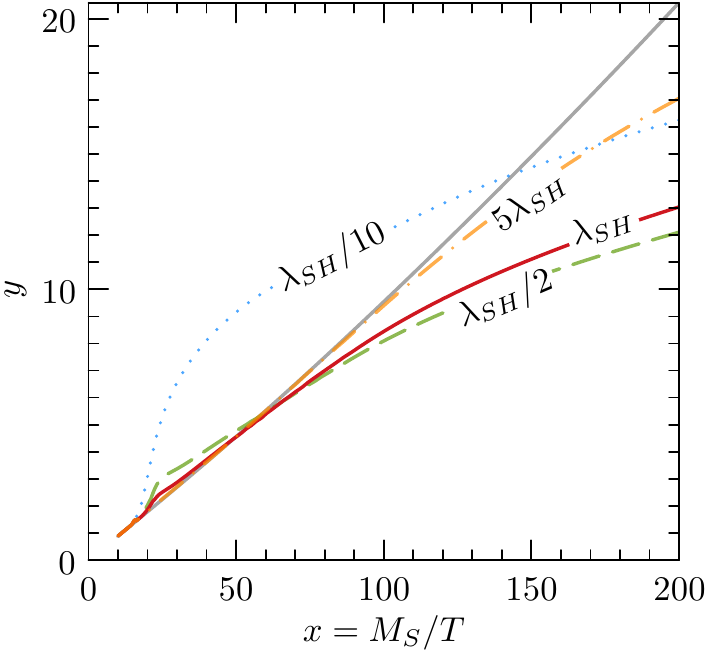}~~~~\includegraphics{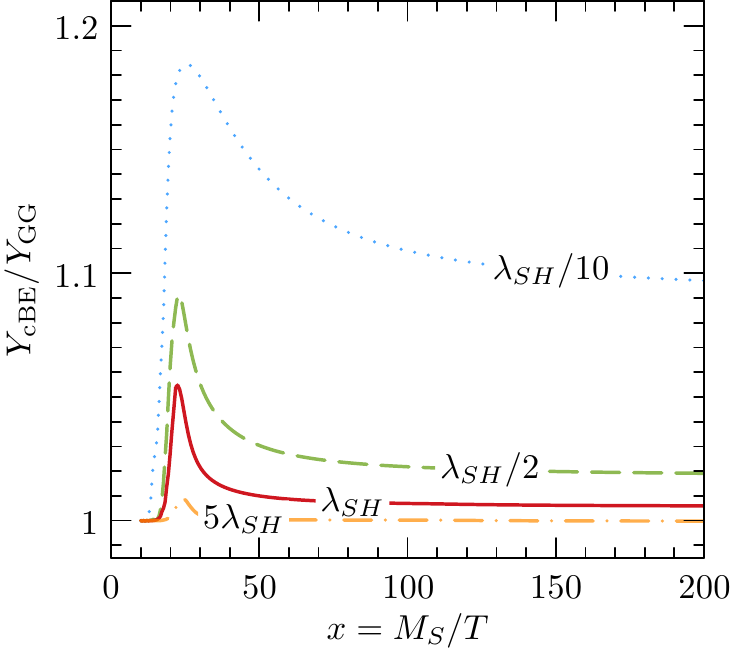}
\caption{The evolution of DM temperature and comoving number density for an example point in the semi-annihilation regime with $M_S=132.5$~GeV. \textit{Left:} a clear departure from kinetic equilibrium (gray solid line) is seen for $\lambda_{SH}=7.08\times 10^{-3}$ that gives the correct relic abundance (red solid), as well as if its value is modified to be 5 times larger (yellow dot-dashed), 2 or 10 times smaller (green dashed and blue dotted, respectively). The bump around $x\approx 20$ is due to `self-heating' caused by semi-annihilation and gets more pronounced, the more suppressed are the elastic scatterings. \textit{Right:} the difference between the cBE and standard Gelmini\&Gondolo \cite{Gondolo:1990dk} treatment for the same set of values of $\lambda_{SH}$. The smaller the Higgs portal coupling, the larger the impact of the final relic abundance.}
\label{fig:evol}
\end{center}
\end{figure}

Let us stress that without implementation of the cBE, one is, in principle, unable to robustly claim that the result of the standard treatment is correct. To illustrate this point, figure~\ref{fig:evol} shows the evolution of the DM temperature and the ratio of the cBE treatment to the standard one for one example point on the absolute stability boundary with mass $M_{S}=132.5$~GeV and $\lambda_{SH}=7.08\times 10^{-3}$. In both panels, the red solid line shows the evolution for this value of the Higgs portal coupling, leading to the correct value of the relic abundance, while additional lines illustrate how the situation would change for values of $\lambda_{SH}$ being 2 or 10 times smaller and 5 times larger. In all of the cases we find kinetic decoupling happening very early indeed, around $x \approx 20$. Moreover, the bump just after this decoupling indicates a period of `self-heating' of the DM component caused by the semi-annihilation. For larger values of $\lambda_{SH}$, this effect is washed out by a still relatively strong elastic scattering, only leading to small deviation from kinetic equilibrium. Nevertheless, it is clear that in all of these cases an assumption of $T_{\rm DM}$ tracing the bath temperature $T$ is incorrect.

The right panel of figure~\ref{fig:evol}, in turn, illustrates the resulting feedback on the number density. As soon as DM starts to leave chemical equilibrium, semi-annihilation heats the bath of DM particles in which the annihilation processes take place to a slightly higher temperature than the bath of photons from which DM is being produced at the same time. This difference in temperatures, through velocity dependence of the cross section, leads to imbalance of the annihilation and production rates. This results in the raise of the ratio $Y_{\rm cBE}/Y_{\rm GG}$, as again seen for all the cases. After the initial period of self-heating, early kinetic decoupling means faster cooling caused by the expansion of the Universe and the reverse happens, with now annihilation being slightly more efficient, resulting in a drop in the above ratio. After freeze-out, the final value of the relic abundance is therefore essentially unaltered, but only due to \textit{a}) a still relatively large coupling $\lambda_{SH}$ needed for obtaining correct relic density and \textit{b}) a velocity dependence of the semi-annihilation cross sections, which is very mild. If either of these conditions were not fulfilled, the effect of early kinetic decoupling on relic abundance would be much more significant.

In summary, we find that in the $\mathbb{Z}_{3}$ singlet DM model the standard treatment of the freeze-out process is a good approximation everywhere apart from the Higgs resonance region, even though also in the semi-annihilation regime the kinetic equilibrium  is not maintained around the time of chemical decoupling. 

%%%%%%%%%%%%%%%%%%%%%%%%%%%%%%%%%%%%%%%%%%%%%%%%%%%%%%%%%%
\section{Direct and indirect detection}
\label{sec:indirect}

\begin{figure}[tb]
\begin{center}
  \includegraphics{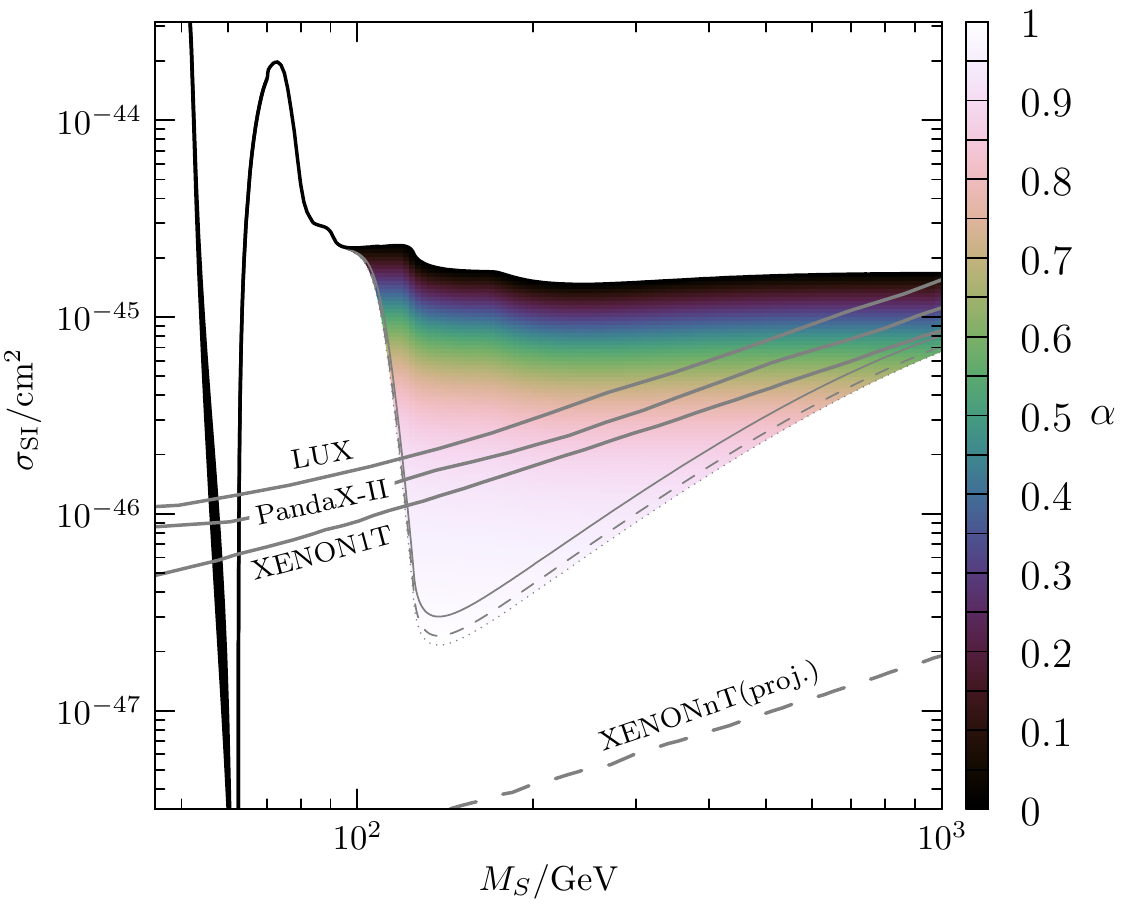}
\caption{Direct detection colour-coded for the semi-annihilation fraction $\alpha$ at the freeze-out. The parameter space is constrained from below by perturbative unitarity (thin solid), the stability of the EW vacuum (thin dashed) and metastability of the vacuum (thin dotted). The experimental bounds from LUX(2017) \cite{Akerib:2016vxi}, PandaX-II (2017) \cite{Cui:2017nnn}, XENON1T(2018) \cite{Aprile:2018dbl} and the projected sensitivity of XENONnT \cite{Ni2017} are shown in grey.}
\label{fig:direct:detection}
\end{center}
\end{figure}

\begin{figure}[tb]
\begin{center}
  \includegraphics{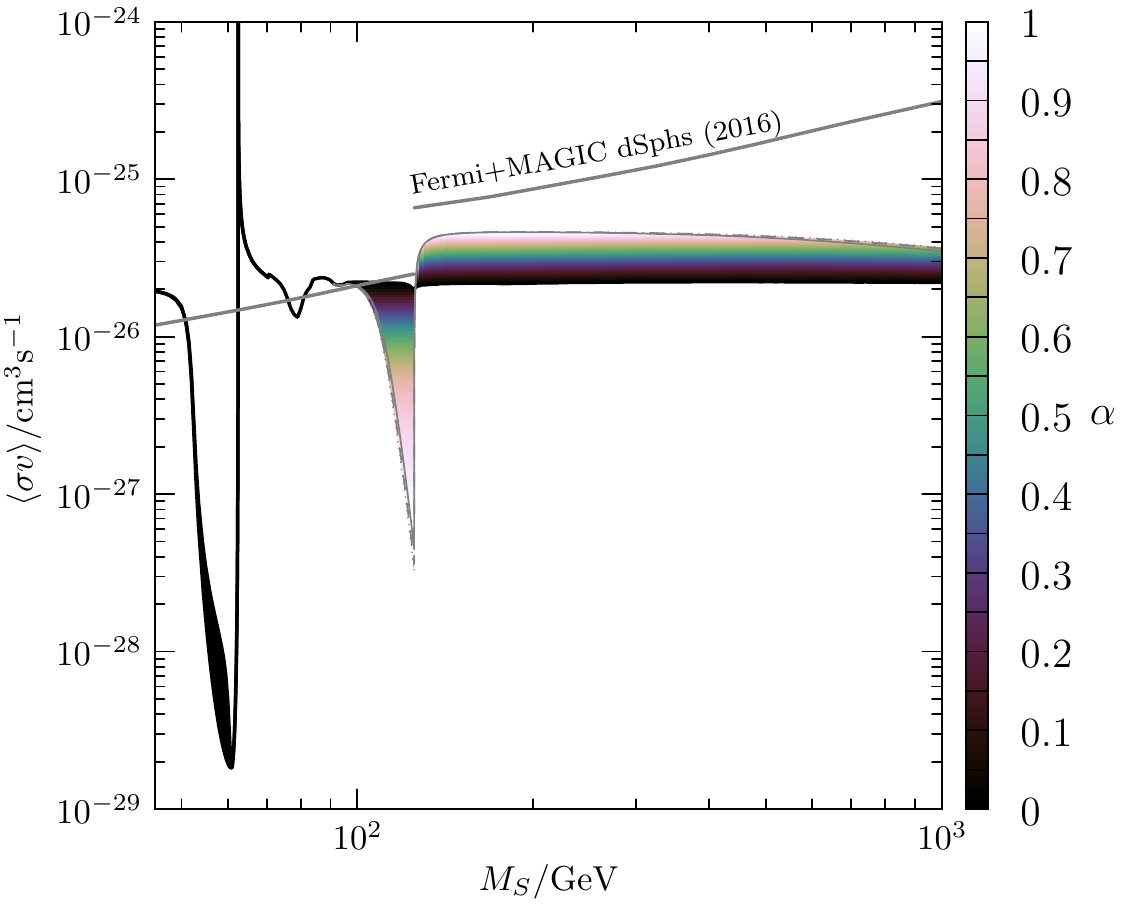}
\caption{The indirect detection signal  colour-coded for the semi-annihilation fraction $\alpha$ at the freeze-out as in figure~\ref{fig:direct:detection}. The observational bound from the combined $\gamma$-ray events from the sky region of the dwarf satellite galaxies (dSphs) measured by the Fermi LAT and MAGIC experiments~\cite{Ahnen:2016qkx}. The sharp step at $M_S \sim M_h$ is due to the fact that the dominant annihilation channel of $S$ switches from the $b\bar b$ channel at below the Higgs mass to the $h h$ channel above $M_{h}$.}
\label{fig:indirect:detection}
\end{center}
\end{figure}

We use the micrOMEGAs \cite{Belanger:2018mqt} to calculate the spin-independent direct detection cross section. The predicted signal and current constraints from LUX(2017) \cite{Akerib:2016vxi}, PandaX-II (2017) \cite{Cui:2017nnn} and XENON1T(2018) \cite{Aprile:2018dbl} are shown in figure~\ref{fig:direct:detection}. The projected sensitivity of the XENONnT \cite{Ni2017} will be sufficient to prove or disprove the current model.

The parameter space is constrained from below by perturbative unitarity (thin solid), the stability of the EW vacuum (thin dashed) and metastability of the vacuum (thin dotted).  The colour code \cite{2011BASI...39..289G} shows the fraction of semi-annihilation defined as
\begin{equation}
\alpha= \frac{\frac{1}{2} \langle v\sigma^{SS\rightarrow S^* h}\rangle}{\langle v \sigma^{SS^* \rightarrow XX}\rangle  + \frac{1}{2} \langle v\sigma^{SS\rightarrow
S^* h} \rangle},
\end{equation}
since $S S \to S^* h$ is the only  semi-annihilation process in this model. 

Around the Higgs resonance, the lower boundary of the allowed area is determined by scenario A and the upper boundary by scenario B of the QCD phase transition (see section \ref{sec:relic}), in which case the cross section is up to $6.7$ times greater than in the standard treatment without taking into account the early kinetic decoupling. For the invisible branching ratio $\text{BR}_{\text{inv}} < 0.17$, the lowest allowed singlet mass is $M_{S} = 54.6$~GeV for scenario A and $M_{S} = 55.4$~GeV for scenario B, due to the up to $6.7$ times greater $\text{BR}_{\text{inv}}$ in that case. Early kinetic decoupling influences relic density mainly for $M_{S}$ below $M_{h}/2$, for which the direct detection cross section can be up to $6.7$ times greater as well. Therefore, from direct detection bounds, the lowest allowed singlet mass is $M_{S} = 56.8$~GeV for scenario A and $M_{S} = 58.4$~GeV for scenario B. With DM mass well below the Higgs mass, semi-annihilation is ineffective and the model predictions for the cross section $\sigma_{\text{SI}}$ coincide with the predictions of the $\mathbb{Z}_{2}$ complex singlet DM model.

At DM masses around and above the Higgs mass, semi-annihilation is important if the singlet cubic self-coupling $\mu_{3}$ is sizeable. At the upper boundary of the $\sigma_{\text{SI}}$ area, $\mu_{3} = 0$ and there is no semi-annihilation. This boundary coincides with the direct detection curve for the usual $\mathbb{Z}_{2}$ complex singlet model. The lower boundary of the allowed region is determined by the maximal allowed $\mu_{3}/M_{S}$ at $\lambda_{S} = \pi$ (the highest value allowed by perturbativity). We show the bounds by perturbative unitarity (thin solid), the stability of the EW vacuum (thin dashed) and metastability of the vacuum (thin dotted), of which unitarity is the strongest constraint.  At high DM masses, the model again approaches the $\mathbb{Z}_{2}$ complex singlet DM. The projected sensitivity of XENONnT \cite{Ni2017} will allow to rule out or discover $\mathbb{Z}_{3}$ dark matter in a large region of the parameter space in the near future.

Figure~\ref{fig:indirect:detection} shows the indirect detection signal and the strongest present constraints. The indirect averaged cross section $\langle \sigma v \rangle$ is calculated with micrOMEGAs with $v \simeq 10^{-3} c$.  The strongest indirect constraints come from the combined data of the Fermi LAT satellite instrument \cite{fermi} and the MAGIC Cherenkov telescope \cite{magic}. The constraints are estimated from the combined $\gamma$-ray events from the sky region of the dwarf satellite galaxies (dSphs)~\cite{Ahnen:2016qkx}.

The colour code denotes the semi-annihilation fraction $\alpha$ at the time of the freeze-out in the same manner as in figure~\ref{fig:direct:detection}. Again, the early kinetic decoupling influences significantly only a small $M_{S}$ region just below $M_{h}/2$, resulting in an up to $6.7$ times greater cross section in scenario B, while in the semi-annihilation region we show that the micrOMEGAs result provides a very good approximation. The sharp dip in $\langle \sigma v \rangle$ below $M_{h}$ results from the drop in temperature after the freeze-out, which makes semi-annihilation processes with $M_{S} < M_{h}$ kinematically forbidden at present times. The Fermi+MAGIC constraint has a sharp step at $M_S \sim M_h$. The reason is that the dominant annihilation channel switches from the $b\bar b$ channel at below the Higgs mass to the $h h$ and $h S$ channels above that mass. Also notice that Ref.~\cite{Ahnen:2016qkx} presents only the constraints for the $WW$ final state, but they are very similar to the $hh$ final state constraints (see e.g.~\cite{Cai:2018imb}). The constraint on the $hS$ final state, however, has to be corrected due to the different boost of the $h$ in this final state. For this, we express $M_S$ from the equation $E_h(M_S) = (3M_S^2 + M_h^2)/(4M_S)$ on the constraint above the Higgs mass and scale the constraint by a factor of two~\cite{Cai:2018imb}. 

Comparing figures~\ref{fig:direct:detection} and~\ref{fig:indirect:detection}, we see that the direct constraints are much more restrictive. In future, new $\gamma$-ray data from Fermi LAT and possibly discovered new dwarfs by the Gaia telescope can strengthen indirect constraints~\cite{Charles:2016pgz}. However, the direct detection constraints are expected to be enhanced even quicker by the XENONnT experiment \cite{Ni2017}.

%%%%%%%%%%%%%%%%%%%%%%%%%%%%%%%%%%%%%%%%%%%%%%%%%%%%%%%%%%
\section{Conclusions}
\label{sec:conclusions}

We consider complex scalar singlet dark matter stabilised by a $\mathbb{Z}_{3}$ symmetry. The presence of a cubic singlet self-coupling considerably changes the DM phenomenology. The cubic coupling gives rise to the semi-annihilation process $S \, S \to S^{*} \, h$, which can dominate the determination of the relic density. Unlike for $\mathbb{Z}_{2}$-symmetric DM models, the relic density and the strength of the direct detection signal are not directly related. The Higgs portal and with it the direct detection cross section can be considerably diminished. However, the cubic coupling cannot be arbitrarily large: it will induce $\mathbb{Z}_{3}$-breaking minima deeper than our minimum, as well as break unitarity.

We calculate the stability and metastability bounds on the one-loop effective potential and use the perturbative unitarity of scattering at finite energy to place a new, robust constraint on the direct detection cross section at large semi-annihilation. These bounds are summarised in figure~\ref{fig:bounds}. 

In addition, we improve the treatment of the thermal freeze-out by including the evolution of the dark matter temperature and its feedback onto relic abundance. This results in a larger Higgs portal -- in particular for dark matter mass below the Higgs resonance. In this regime, the larger than expected portal coupling leads to stronger constraints from the Higgs invisible width and direct detection signals. The Higgs invisible width and direct and indirect detection signals can be up to $6.7$ times greater than for the standard result. The updated direct detection constraints are shown in figure~\ref{fig:direct:detection} and the indirect detection constraints in figure~\ref{fig:indirect:detection}.

In contrast, in the strong semi-annihilation regime, the results do not differ from the standard treatment by more than a couple of per cent. Nonetheless, this is not given \textit{a priori}, considering that the temperature of DM substantially differs from the equilibrium, as shown in figure~\ref{fig:evol}. It would be interesting to study this effect in other models with early kinetic decoupling, such as models that also include an inert doublet \cite{Belanger:2012vp,Belanger:2014bga}.

The presently allowed mass ranges for the model are $56.8~\text{GeV}$ to $~58.4~\text{GeV}$ (depending on the details of early kinetic decoupling) $\lesssim  M_{S} \lesssim 62.8~\text{GeV}$ around the Higgs resonance and at $M_{S} \gtrsim 122~\text{GeV}$ in the region with semi-annihilation. The constraint from perturbative unitarity comes close to excluding the model near and above TeV-scale. New results from the XENONnT experiment will be sufficient to prove or rule out the model in the semi-annihilation region. At higher scales, however, the model becomes indistinguishable from the usual $\mathbb{Z}_{2}$ dark matter and is constrained only by perturbativity and unitarity of the Higgs portal coupling. In the narrow allowed range with $M_{S} < M_{h}/2$, the Higgs invisible branching ratio could be measured by the LHC or a future electron-positron collider.

The singlet self-coupling must be at least $\lambda_{S} \gtrsim 0.2\, \pi$ in order to not forbid the large semi-annihilation region by the direct detection results of XENON1T. Because of that, the model cannot remain valid up to the Planck scale. Discovery of $\mathbb{Z}_{3}$ scalar singlet dark matter may therefore entail a relatively low new physics scale with new fermions or a dark $U(1)$ symmetry.

%%%%%%%%%%%%%%%%%%%%%%%%%%%%%%%%%%%%%%%%%%%%%%%%%%%%%%%%%%
\acknowledgments

We would like to thank Genevi\`{e}ve B\'{e}langer, Alexander Pukhov, Jan Conrad, Torsten Bringmann, and Carlo Marzo for discussions. This work was supported by the Estonian Research Council grant PRG434, the grant IUT23-6 of the Estonian Ministry of Education and Research, and by the EU through the ERDF CoE program project TK133. At the University of Oslo, A. Hryczuk was supported by the Strategic Dark Matter Initiative (SDI).

%%%%%%%%%%%%%%%%%%%%%%%%%%%%%%%%%%%%%%%%%%%%%%%%%%%%%%%%%%
\appendix
\section{Field dependent masses and counter-terms}
\label{sec:counter}

We decompose the singlet as
\begin{equation}
  S = s e^{i \phi_{S}} = \frac{S_{R} + i S_{I}}{\sqrt{2}}.
\end{equation}

It is convenient to minimise the potential in the first parameterisation: because we have chosen $\mu_{3} \geqslant 0$, we have $\cos \phi_{S} = -1$ in minima of potential with $\langle s \rangle \equiv v_{s} \neq 0$. We have $\langle S_{R} \rangle = \sqrt{2} v_{s}$.

The field-dependent masses of the Higgs boson and the real component of $S$ are given by the eigenvalues $m^{2}_{1,2}$ of the mass matrix
\begin{equation}
  m_{R}^{2} = 
  \begin{pmatrix}
    \mu_{H}^{2} + 3 h^{2} \lambda_{H}+ \frac{1}{2} \lambda_{SH} S_{R}^2 & \lambda_{SH} h S_{R}
    \\
    \lambda_{SH} h S_{R} & \mu_{S}^{2} + 3 \lambda_{S} S_{R}^2 + \frac{3}{2} \sqrt{2} \mu_{3} S_{R} 
     + \frac{1}{2} h^{2} \lambda_{SH}
  \end{pmatrix}.
\end{equation}
The masses and degrees of freedom $n_{i}$ of the fields are given in table~\ref{tab:field:dependent:masses}. We neglect the contributions of the Goldstone bosons $G^{0}$ and $G^{\pm}$.

We can use the first six counter-terms in the counter-term potential \eqref{eq:delta:V:Z:3:singlet} to fix some VEVs of the fields and masses to their tree-level values. The overall constant counter-term $\delta V_{0}$ can be used to fix vacuum energy to the measured level.

Some values do not change from tree-level as it is. In particular, the VEV of the imaginary part of $S$ does not change with quantum corrections, once it is zero:
\begin{equation}
  \left. \partial_{S_{I}} \Delta V \right|_{S_{I} = 0} = 0,
\end{equation}
so we can restrict our attention to minima with $S_{I} = 0$ without loss of generality (in the case of broken $\mathbb{Z}_{3}$, we can consider that of the three degenerate minima for which $S_{I} = 0$).
Once $S_{I} = 0$, the VEV for $S_{R}$ does not shift from the tree-level minimum either:
\begin{equation}
  \left. \partial_{S_{R}} \Delta V \right|_{S_{R} = S_{I} = 0} = 0.
\end{equation}
If $S_{I} = 0$, then only the mass mixing term $m_{h S_{R}}^{2}$ for real scalars can potentially be non-zero. If $S_{R} = 0$, it is zero at tree-level and stays zero at loop-level due to unbroken $\mathbb{Z}_{3}$:
\begin{equation}
  \left. \partial_{h} \partial_{S_{R}} \Delta V 
  \right|_{S_{R} = S_{I} = 0} = 0.
\end{equation}

\begin{table}[tb]
\caption{Field-dependent masses and degrees of freedom.}
\begin{center}
\begin{tabular}{ccc}
  Field $i$ & $m_{i}^{2}$ & $n_{i}$ \\
  \hline
  $h$ & $m^{2}_{1}$ & $1$ \\
  $S_{R}$ & $m^{2}_{2}$ & $1$ \\
  $S_{I}$ & $\mu_{S}^{2} + \lambda_{S} S_{R}^2 - \frac{3}{\sqrt{2}} \mu_{3} S_{R} 
  + \frac{1}{2} \lambda_{SH} h^{2}  $ & $1$ \\
%  $G^{0}$ & & $1$ \\
%  $G^{+}$ & & $1$ \\
  $Z^{0}$ & $\frac{1}{4} (g^{2} + g^{\prime 2}) h^{2}$ & $3$ \\
  $W^{\pm}$ & $\frac{1}{4} g^{2} h^{2}$ & $6$ \\
  $t$ & $\frac{1}{2} y_{t} h^{2}$ & $-12$ \\
\end{tabular}
\end{center}
\label{tab:field:dependent:masses}
\end{table}

The loop corrections to the Higgs portal $\lambda_{SH}$ are proportional to $\lambda_{SH}$ itself and therefore negligible for the relatively small values of the portal that yield the correct relic density.

The renormalisation conditions that we do need are
\begin{equation}
  \left. \partial_{h}(\Delta V + \delta V) \right|_{h = v, \, S_{R} = 0} = 0,
  \label{eq:Higgs:VEV:EW:CT}
\end{equation}
to ensure that the Higgs VEV does not move from its tree-level position in our minimum,
\begin{equation}
  \left. \partial_{h}(\Delta V + \delta V) \right|_{h = v_{h}, \, S_{R} = \sqrt{2} v_{s}} = 0, 
  \quad
  \left. \partial_{h}(\Delta V + \delta V) \right|_{h = v_{h}, \, S_{R} = \sqrt{2} v_{s}} = 0,
   \label{eq:VEVs:other:CT}
\end{equation}
to ensure that the VEVs do not move from their tree-level positions in the other minimum with $h = v_{h}$ and $S_{R} = \sqrt{2} v_{s}$,
\begin{equation}
  \left. \partial_{h}^{2} (\Delta V + \delta V) \right|_{h = v, \, S_{R} = 0} = 0,
  \quad
  \left. \partial_{S_{R}}^{2} (\Delta V + \delta V) \right|_{h = v, \, S_{R} = 0} = 0,
  \label{eq:masses:EW:CT}
\end{equation}
to keep the mass matrix in our minimum from changing, and
\begin{equation}
    \left. \partial_{S_{R}}^{3} (\Delta V + \delta V) \right|_{h = v, \, S_{R} = 0} = 0
  \label{eq:cubic:EW:CT}
\end{equation}
to keep the singlet cubic coupling at its tree-level value in our minimum.

Solving the system of equations \eqref{eq:Higgs:VEV:EW:CT}, \eqref{eq:VEVs:other:CT}, \eqref{eq:masses:EW:CT} and \eqref{eq:cubic:EW:CT}, we obtain for the counter-terms
\begin{align}
  \delta \mu_{H}^{2} &= \frac{1}{2 v} (v \partial_{h}^{2} \Delta V|_{h = v, \, S_{R} = 0} - 3 \partial_{h} \Delta V|_{h = v, \, S_{R} = 0}),
  \\
  \delta \lambda_{H} &= \frac{1}{2 v^{3}} (\partial_{h} \Delta V|_{h = v, \, S_{R} = 0} - v \partial_{h}^{2} \Delta V|_{h = v, \, S_{R} = 0}),
  \\
  \delta \mu_{S}^{2} &= \frac{1}{4 v v_{h} v_{s}^{2}} 
  [v_{h} (v_{h}^{2} - 3 v^{2}) \partial_{h} \Delta V|_{h = v, \, S_{R} = 0} -4 v v_{h} v_{s}^{2} \partial_{S_{R}}^{2} \Delta V|_{h = v, \, S_{R} = 0} 
  \notag
  \\
  &+ 2 v^{3} \partial_{h} \Delta V|_{h = v_{h}, S_{R} = \sqrt{2} v_{s}} + v v_{h} (v^{2} - v_{h}^{2}) \partial_{h}^{2} \Delta V|_{h = v, \, S_{R} = 0}],
  \\
  \delta \lambda_{S} &= \frac{1}{8 v^{3} v_{h} v_{s}^{4}} [-2 \sqrt{2} v^{3} v_{h} v_{s} \partial_{S_{R}} \Delta V|_{h=v_{h},\,S_{R} = \sqrt{2} v_{s}} 
  \notag
  \\
  &+ 4 v^{3} v_{h} v_{s}^{2} \partial_{S_{R}}^{2} \Delta V|_{h=v,\, S_{R} = 0}
  + 2 \sqrt{2} v^{3} v_{h} v_{s}^{3} \partial_{S_{R}}^{3} \Delta V|_{h=v,\,S_{R} = 0}
  \notag
  \\
  & + 3 v^{4} v_{h} \partial_{h} \Delta V|_{h=v,\, S_{R} = 0} - 4 v^{2} v_{h}^{3} \partial_{h} \Delta V|_{h=v,\, S_{R} = 0}
  \notag
  \\
  & + v_{h}^{5} \partial_{h} \Delta V|_{h=v,\, S_{R} = 0} - 2  v^{5} \partial_{h} \Delta V|_{h=v,\, S_{R} = \sqrt{2} v_{s}} 
  \notag
  \\
  & + 2 v^{3} v_{h}^{2} \partial_{h} \Delta V|_{h=v,\, S_{R} = \sqrt{2} v_{s}} 
  - v v_{h} (v^{2} - v_{h}^{2})^{2} \partial_{h}^{2} \Delta V|_{h=v,\, S_{R} = 0}],
  \\ 
  \delta \lambda_{SH} &= \frac{1}{2 v^{3} v_{h} v_{s}^{2}} [v_{h} (3 v^{2} - v_{h}^{2}) \partial_{h} \Delta V|_{h=v, \, S_{R} = 0} - 2 v^{3} \partial_{h} \Delta V|_{h = v_{h}, \, S_{R} = \sqrt{2} v_{s}} 
  \notag
  \\
  &+ v v_{h} (v_{h}^{2} - v^{2}) \partial_{h}^{2}|_{h=v,\,S_{R} = 0}],
  \\
  \delta \mu_{3} &= -\frac{\sqrt{2}}{3} \partial_{S_{R}}^{3} \Delta V|_{h=v, \, S_{R} = 0}.
\end{align}

%%%%%%%%%%%%%%%%%%%%%%%%%%%%%%%%%%%%%%%%%%%%%%%%%%%%%%%%%%
\bibliography{Z3SDMrev}
\end{document}